# Interlayer bonding investigation of 3D printing cementitious materials with fluidity−retaining polycarboxylate superplasticizer and high-dispersion polycarboxylate superplasticizer

Tinghong Pan[a, b], Yaqing Jiang[a*], Xuping Ji[a, c*]

[a] *College of Mechanics and Materials, Hohai University, Nanjing 211100, China;*

[b] *Yunnan Provincial Key Laboratory of Civil Engineering Disaster Prevention, Faculty of Civil Engineering and Mechanics, Kunming University of Science and Technology, Jingming South Road, Kunming, 650500, China;*

[c] *State Key Laboratory of High Performance Civil Engineering Materials, Jiangsu Sobute New Materials Co. Ltd., Nanjing, 211103, China;*

[*]*corresponding author*

*E−mail addresses: yqjiang@hhu.edu.cn (Y. Jiang), xpji@hhu.edu.cn (X. Ji).*

**Abstract:**

Proposed special requirements exist for the rheological properties and time-varying characteristics of 3D printing cementitious materials (3DPC). In this study, high-dispersion polycarboxylate superplasticizer (HD-PC) and fluidity−retaining polycarboxylate superplasticizer (FR-PC) were used to control the rheological behaviors of 3DPC. The correlation of the time interval, the time varying characteristics of the rheological properties and the interlayer bonding strength were investigated. The results indicated that FR-PC improved the fluidity-retention ability and thixotropy of the fresh pastes. The thixotropic hysteresis loop area and reflocculation rate ($R_{thix}$) of FR-PC were 101.9% and 80.4% higher than those of HD-PC, respectively. Furthermore, the FR-PC polymer has a positive effect on the interlayer bonding and may reduce the negative effect caused by extending the time interval. During the time interval of 20 s to 30 min, the interlayer bonding strength with FR-PC decreases by 14.1%, while that with HD-PC decreases by 50.0%.

---

[*] Corresponding authors.
E-mail addresses: yqjiang@hhu.edu.cn (Y. Jiang), xpji@hhu.edu.cn (X. Ji).

**Keywords:** Cementitious materials; 3D printing technology; Rheological properties; Interlayer bonding strength; X–ray micro computed tomography

## 1. Introduction

Intelligent construction is an important development direction of the construction industry in the future. 3D printing cementitious materials (3DPC) is the core and key to realizing intelligent building manufacturing [1-3]. For the construction industry, 3DPC breaks through the limitations of the traditional formwork−casting method and may help to build structures with complex shapes that cannot be built by the traditional formwork−casting method. Compared with the traditional formwork−casting method, 3DPC brings significant technological advancements for the building−construction industry, such as no required formwork [4-5], reduced execution period labor costs [1,6-7], reduced waste and increased freeform architectural designs [6,8].

The rheological behavior and thixotropic behavior of 3DPC plays a key role in the printing process [9-10]. Different rheological properties are required during and after extrusion [11-12]. During the pumping and extrusion process, 3DPC should have higher workability, lower yield stress and lower viscosity for easy pumping and extrusion [3]. However, excessive workability may reduce the bearing capacity of the subsequent deposited layer and fail to meet the shape stability criteria of the printed structures. After extrusion, the mixtures need high static yield stress and high viscosity to withstand the weight of the upper layers and to resist structural deformation [13-15].

In addition, the rheological and thixotropic properties of 3DPC also affect the interlayer bonding between layers. The mixture needs good wettability and low dynamic yield stress to automatically adjust its shape under self-weight to match the rough surface of the bottom layer. Furthermore, these properties help to successfully fill and permeate the pores and cracks in the surface of the previously deposited layer, thus leading to the initial formation of mechanical interlocking between the two layers [16-18]. However, if the dynamic yield stress of the mixtures is too large, part of the air will be trapped between the two layers, which will reduce the contact area between adjacent layers, thus affecting the interlayer bonding strength of the printed structures [11].

In addition, the printing time interval, as one of the critical factors in 3DPC, should be considered. Extending the printing time interval would lead to a decrease in the interlayer bonding strength. This may be attributed to the water loss occurs between the two layers [19-21]. The evaporation and drying of water on the interlayer surface during the printing interval will cause insufficient hydration and high porosity, resulting in a decrease in interlayer bond strength [19,22]. Before the layers contact, the bottom printed layer has a short time to rest and build the structure. As reported by Roussel et al. [23], there is a critical delay between two layers. If the printing time interval is above the critical delay, the bottom printed layer builds too much structure, and its static yield stress increases above a critical value. The stresses generated by the flow of the upper layer and movement of the nozzle are not sufficient to initiate flow in the resting bottom layer [18,23]. Thus, the two layers do not mix at all, as vibrating is prohibited in the case of 3DPC, and a weak mechanical consolidation between two layers may appear in the final structure.

As noted above, reducing the printing time interval may be a way to improve the interlayer bonding performance [24-25]. However, the printing time interval is affected by the geometric shape of the printing structures and the printing speed [26]. A more complex shape of the printed structure will increase the printing time interval [27]. Providing high and sustained flow characteristics for mixtures to ensure enough workability and as low a yield stress as possible until the layers contact may also improve the interlayer bonding performance. However, reports about the variation of rheological behaviors (i.e., static yield stress and fluidity) of fresh pastes with different time intervals are few, and the study of the correlation between the time-varying characteristics of rheological behaviors and the interlayer bonding performance has not been found.

In this paper, two types of PC (high-dispersion polycarboxylate superplasticizer HD-PC and fluidity−retaining polycarboxylate superplasticizer FR-PC) were prepared. Their structural characterization and adsorption behaviors were investigated to reveal the mechanism of action of PC. The rheological behaviors (i.e., static yield stress and fluidity) with different resting times were investigated to reveal the time-varying characteristics of the rheological properties. Additionally, the microstructure and interlayer bonding strength of printed structures with different types of PC (HD-PC and FR-PC) and different printing time intervals (20 s, 15 min, 30 min, 45 min and 60 min) were investigated to explore the correlation of printing time interval, time-varying characteristics of rheological properties and interlayer bonding strength.

## 2. Materials & methods

*2.1 Raw materials*

Ordinary Portland cement (OPC, Type II, 42.5 grade, Nanjing Conch Cement Co. Ltd, China) was selected as the binder. Attapulgite clay (Jiangsu Jiuchuan Nano−material Technology Co., Ltd.), exfoliated into nanoparticles (called nano clay, Nc) shaped with 135 nm average length and approximately 58 nm diameter, was selected as the thixotropic admixture [28]. The chemical compositions of OPC and Nc are given in Table 1. Hydroxypropyl Methylcellulose (HPMC, Renqiu Cheng Yi Chemical Co.Ltd.) was selected as the viscosity modifier.

Table 1. Chemical compositions of OPC (Type II) and Nc [wt.%].

| Materials | CaO | SiO$_2$ | Al$_2$O$_3$ | FeO$_3$ | Na$_2$O | MgO | K$_2$O | SO$_3$ | TiO$_2$ | L.O.I |
|---|---|---|---|---|---|---|---|---|---|---|
| OPC | 62.60 | 21.65 | 5.56 | 4.32 | 0.24 | 0.84 | 0.76 | 2.85 | − | 1.27 |
| Nc | 9.62 | 58.4 | 26.73 | 0.51 | 0.21 | 0.20 | 3.05 | − | 0.15 | 1.13 |

Two types of superplasticizers, commercial high-dispersion polycarboxylate superplasticizer (HD-PC, Jiangsu Sobute New Materials Co. Ltd, Nanjing, China), and self-synthesized fluidity−retaining polycarboxylate superplasticizer (FR-PC) were employed as rheological modifiers. FR-PC polymer was prepared via semibatch free radical polymerization at 80 °C by copolymerizing the monomer of acrylic acid (AA), sodium allyl sulfonate (SAS), polyoxyvinyl unsaturated ester macromonomer (MPEG400MA) and α-methallyl-ω-hydroxy poly (ethylene glycol) ether macromonomer with Mw of ca. 2400 (HPEG2400). These monomers were mixed with a monomer molar ratio of 3 (AA):1 (SAS):0.3 (MPEG400MA):0.7 (HPEG2400). Ammonium persulfate was used as the initiator, and its dosage was 40% of the total mass of monomer.

*2.2 Characterization of Pc polymers*

*2.2.1 FTIR analysis*

The chemical structures of the HD-PC polymer and FR-PC were characterized by an FTIR (Thermo Scientific Nicolet iS5, USA) instrument. The PC solution was homogeneously mixed with KBr salt and completely dried in a vacuum oven at 80 °C. The mixture was pressed into a tablet and then scanned in transmission mode from 400 to 4000 cm$^{-1}$ at a room temperature of 25 °C.

*2.2.2 Gel permeation chromatography*

The weight−average molecular weight (Mw), number−average molecular weight (Mn) and polydispersity index (PD: Mw/Mn) of the PCs were determined from gel permeation chromatography (GPC) measurements using a gel permeation chromatograph (GPC) apparatus (PL−GPC50; PL−GPC220; Waters GPC 1515). In this measurement, 0.1 mol/L Na$_2$SO$_4$ aqueous solution was used as the eluant at a flow rate of 0.5 mL/min. Monodispersive sodium polyethylene sulfonate was used as the standard phase of calibration.

*2.2.3 Adsorption behaviors of PC polymers*

Solutions (20 mL) with different types of PC (HD-PC polymer and FR-PC polymer) and different concentrations (0.2, 0.4, 0.8, 1.2 and 1.6 g/L) were prepared. Cement (5.0 g) was mixed with the solutions by stirring for 10 min and 60 min, respectively. Then, a centrifugation instrument was used to separate these suspensions for 5 min at 4000 r/min. Then, a TOC analyzer (ET1020A Total organic Carbon, Shanghai Euro Tech Ltd, China) was used to measure the organic carbon content in the centrifuged solutions. The concentration of PCs in the centrifuged solutions could be calculated according to the TOC results. The adsorption amount of PC ($A$, mg/g-cement) could be calculated according to the formula below:

$$A = V(C_o - C)/m \qquad (1)$$

where $A$ is the adsorption amount of PC (mg/g-cement); $C_0$ is the initial concentration (g/L); $C$ is the concentration (g/L) of the centrifuged solution after adsorption; $V$ is the solution volume (mL); $m$ is the mass of the cement (g).

*2.2.4 Fluidity*

The fluidity of cement pastes with HD-PC or FR-PC was determined by the mini-slump test. The mass solid/solid ratios of PC to cement (P/C) were fixed at 0.12%, and the mass ratio of water to cement in all the cement pastes was fixed at 0.29. The cement paste specimen was sealed in a container for 0, 15, 30, 45 and 60 min before the measurement. Then, the prepared fresh paste was poured into a mini-slump cone (a height of 60 mm, upper diameter of 36 mm and a bottom diameter of 60 mm) and the cone was quickly lifted up. After the paste stopped flowing, the average value of the maximum diameter and width perpendicular was recorded. The fluidity measurement was repeated 3 times for each sample, and an average of these measured values was considered to be the final fluidity value. The decrease in fluidity with time was used to evaluate the fluidity-retention of the cement paste.

*2.3 Preparation of printable cement pastes*

Two printable mixtures developed by the authors [34] were used in this study, as shown in Table 2. Except for the type of superplasticizer, both mixture compositions were kept identical. The mass ratios of cementitious materials to sand were fixed at 1:1.5, and the mass ratios of water to cementitious materials were fixed at 0.32:1. The dosages of Nc, HPMC and PCs were fixed as 0.8%, 0.24% and 0.3% of the cementitious materials (by mass), respectively.

Table 2. Proportion of 3D printing cement mortars.

| Material | Cement | Quartz sand | Water | Nano−clay | HPMC | FR-PC or HD-PC polymer |
|---|---|---|---|---|---|---|
| Quantity(g) | 1400 | 2100 | 460 | 11.2 | 3.5 | 4.2 |

A JJ−5 planetary cement mortar mixer was used to mix the materials. First, dry batches, such as cement, fine aggregate and nanoclay, were mechanically mixed for 60 s at a slow speed of 140 rpm to form a homogenous mixture. Next, the water and superplasticizer were poured into the mixer and mechanically mixed with a dry batch for 60 s at a slow speed of 140 rpm. Then, the mixture was stopped for 30 s to scrape off the residual slurry on the wall of the mixer. Last, the cement paste was mechanically mixed for 90 s at a high speed of 285 rpm. The mixture was immediately placed into a specific container for rheological property test and poured into a 3D printer.

*2.4 Rheological properties of printable cement pastes*

Rheological properties, with yield stress and plastic viscosity as two intrinsic physical parameters, are of great importance in describing the pumpability, extrudability and buildability of cementitious materials [12,29]. In this study, the rheological properties of fresh pastes, such as viscosity, yield stress, static yield stress and the thixotropic hysteresis loop area were characterized using a BROOKFIELD RST−SST rheometer. Two rheological testing protocols were proposed to quantitatively measure the thixotropic hysteresis loop area and static yield stress.

*2.4.1 Thixotropic hysteresis loop*

The thixotropic hysteresis loop method is one of the most popular methods to evaluate the thixotropy of cement mortar [30]. As shown in Fig. 1, fresh paste is first pre-sheared for 120 s by applying a shear rate sweep from 0 s$^{-1}$ to 100 s$^{-1}$, which is mainly used to create a uniform environment with little test error. Then, after 10 s of rest, an increasing shear rate ramp from 0 s$^{-1}$ to 100 s$^{-1}$ within 100 s is applied to produce the up−curve of the flow test. Finally, the shear rate decreases from 100 s$^{-1}$ to 0 s$^{-1}$ within 100 s to obtain the down−curve of the flow test. In the up−curve of the thixotropic hysteresis loop testing protocols, there are more flocculation structures in the cement suspension, which increases the shear resistance of the rotor, and results in a higher shear stress [31]. In the down−curve of the thixotropic hysteresis loop testing protocols, part of the flocculation structure is destroyed, which decreases the shear resistance of the rotor and results in a lower shear stress than that in the up−curve [31]. The hysteresis loop area between the up−curve and down−curve has the physical dimension of energy per unit time and unit volume [32]. A greater hysteresis loop area implies a higher degree of thixotropy. Thixotropy is assessed as the area enclosed between 20 s$^{-1}$ and 80 s$^{-1}$ strain rates in a hysteresis loop, because the shear stresses tend to be more stable in this region.

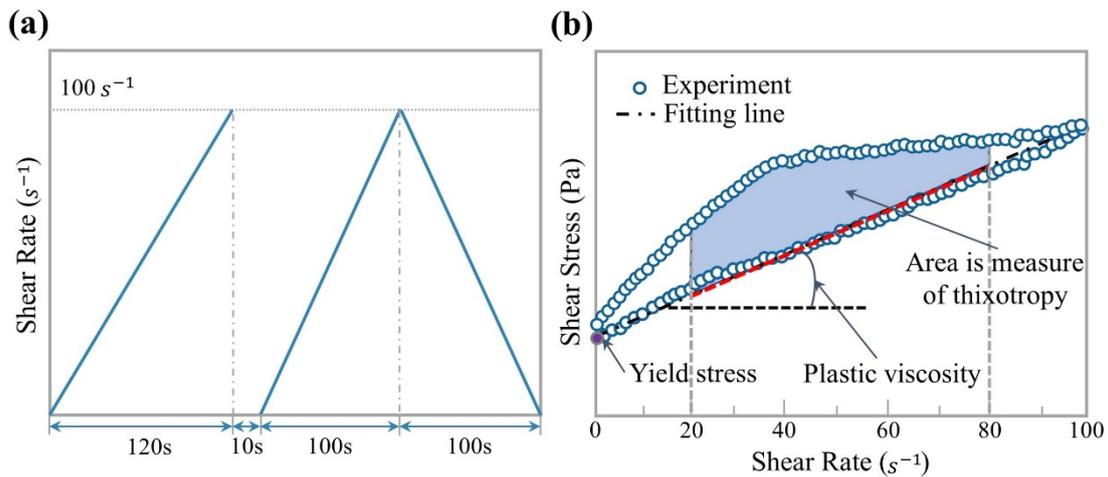

Fig. 1. Rheological testing protocols for thixotropic hysteresis loop.

*2.4.2 Static yield stress*

This protocol is used to monitor the development of static yield stress with resting time. After the fresh paste is evenly mixed via a JJ−5 planetary cement mortar mixer, it is immediately poured into a cylindrical container for rheological property testing. First, fresh paste is pre−sheared for 120 s by applying a shear rate sweep from 0 $s^{-1}$ to 100 $s^{-1}$, which is mainly used to ensure the same initial conditions as other samples. Next, the mixture will rest for a short time (0 s, 150 s, 300 s, 450 s, 600 s, 750 s, 900 s, 1050 s, 2000 s and 3000 s) to build the structure. Then, a content shear rate of 0.02 $s^{-1}$ is applied to break the structure build in the rest time. In this stage, the curve of shear stress vs. time is shown in Fig. 2. The maximum shear stress on the curve is defined as the static yield stress [33]. According to the research results of Perrot and Weng et al. [14, 34], the development of static yield stress can be used to monitor the structural build-up of cement-based materials.

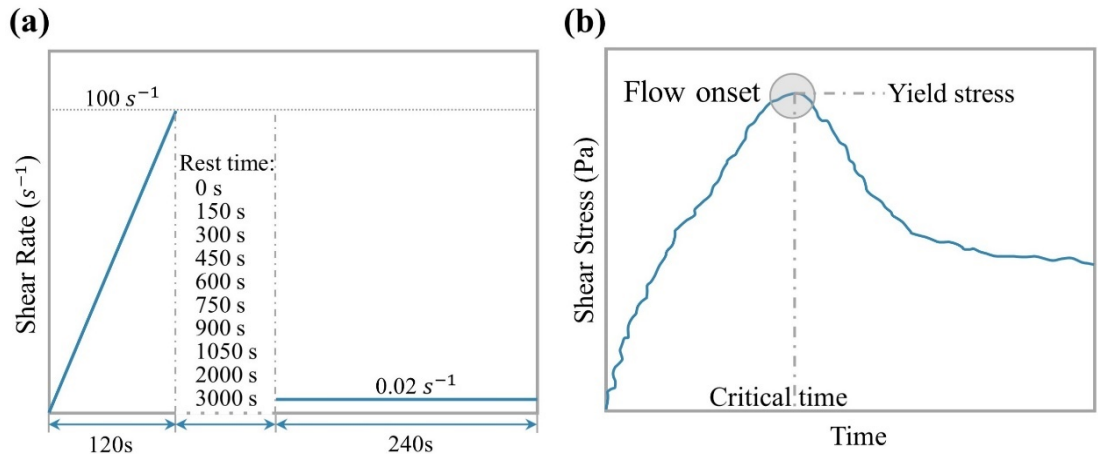

Fig. 2. Test protocols of static yield stress (a), and static yield stress obtained from the peak value (b).

*2.5 Printing device and printing parameters*

A homemade laboratory level gantry 3D printer with forming space of 500 mm × 500 mm × 500 mm was used with a screw extrusion, as shown in Fig. 3. A rectangular nozzle with an opening of 40 mm × 20 mm was used in this work, which results in a large surface contact area between the adjacent extruded layers. The moving speed *v* of the nozzle was set as 60 mm/s, and the height *h* between the nozzle and the build surface was fixed at 25 mm. The screw rod driven by the stepper motor was used to extrude the 3D printed ink out of a nozzle, and the extrusion rate *E* was set as 12 mL/s, by adjusting the rotation rate of the screw rod.

Double−layer structures with different types of superplasticizers and different printing time intervals were prepared for the interlayer bonding strength test and X−CT test, as shown in Table 3. After the double−layer structure was printed, the strip sample was immediately cut into small squares with a length of 40 mm. Then, the cut specimen was cured in the standard curing environment for 28 days.

Table 3. Samples with different types of superplasticizers and different printing time intervals.

| No. | Printing time interval (s) | Types of superplasticizers |
|---|---|---|
| HD-PC/T20 | 20 | HD-PC |
| HD-PC/T900 | 900 | HD-PC |
| HD-PC/T1800 | 1800 | HD-PC |
| HD-PC/T2700 | 2700 | HD-PC |
| HD-PC/T3600 | 3600 | HD-PC |
| FR-PC/T20 | 20 | FR-PC |
| FR-PC/T900 | 900 | FR-PC |
| FR-PC/T1800 | 1800 | FR-PC |
| FR-PC/T2700 | 2700 | FR-PC |
| FR-PC/T3600 | 3600 | FR-PC |

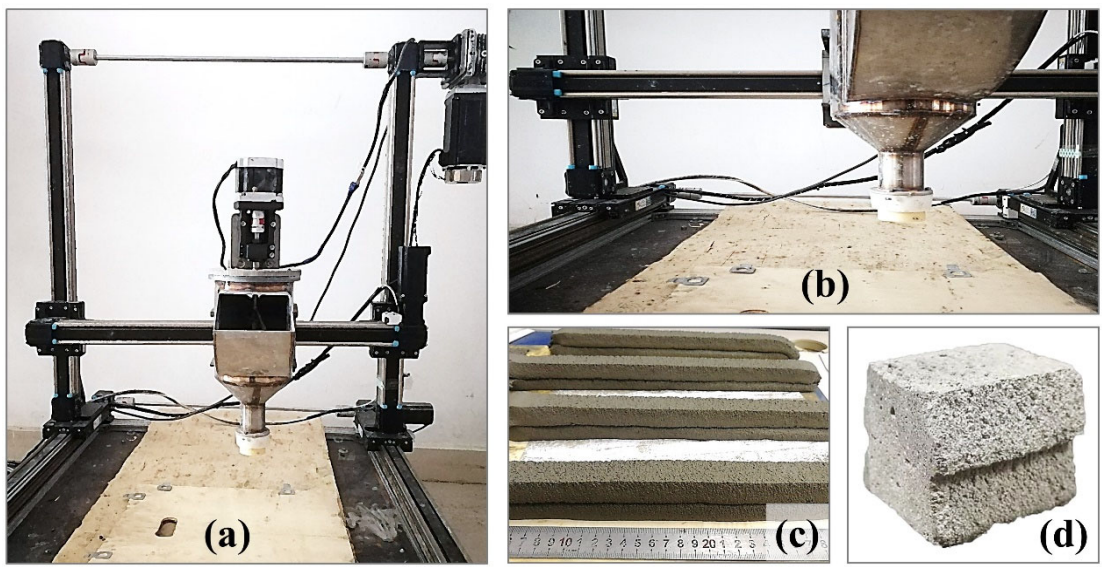

Fig. 3. 3D Printing device (a), printing nozzle (b), double−layer structures without cutting (c) and with cutting (d).

*2.6 Microstructure of the printed structure*

*2.6.1 X-CT image acquisition and segmentation*

The microstructure of the printed structures was determined using a Nikon metrology X−ray CT computed tomography (Nikon XTH) system supplied from Nikon Metrology Instruments.

*2.6.2 Air void characterization*

In this work, the plane porosity and fractal dimension of a 2D slice perpendicular to the stacking direction of the printed layer were used to quantitatively characterize the pore characteristics of 3D printed cement mortars. Porosity can directly reflect the compactness of the printed structures and can be obtained from the proportion of pore area in the total area of the 2D slice.

$$\phi = \frac{A_V}{A_T} \qquad (2)$$

where $\phi$ is the plane porosity, $A_V$ is the area of void space in the 2D slice, and $A_T$ is the total area of the 2D slice.

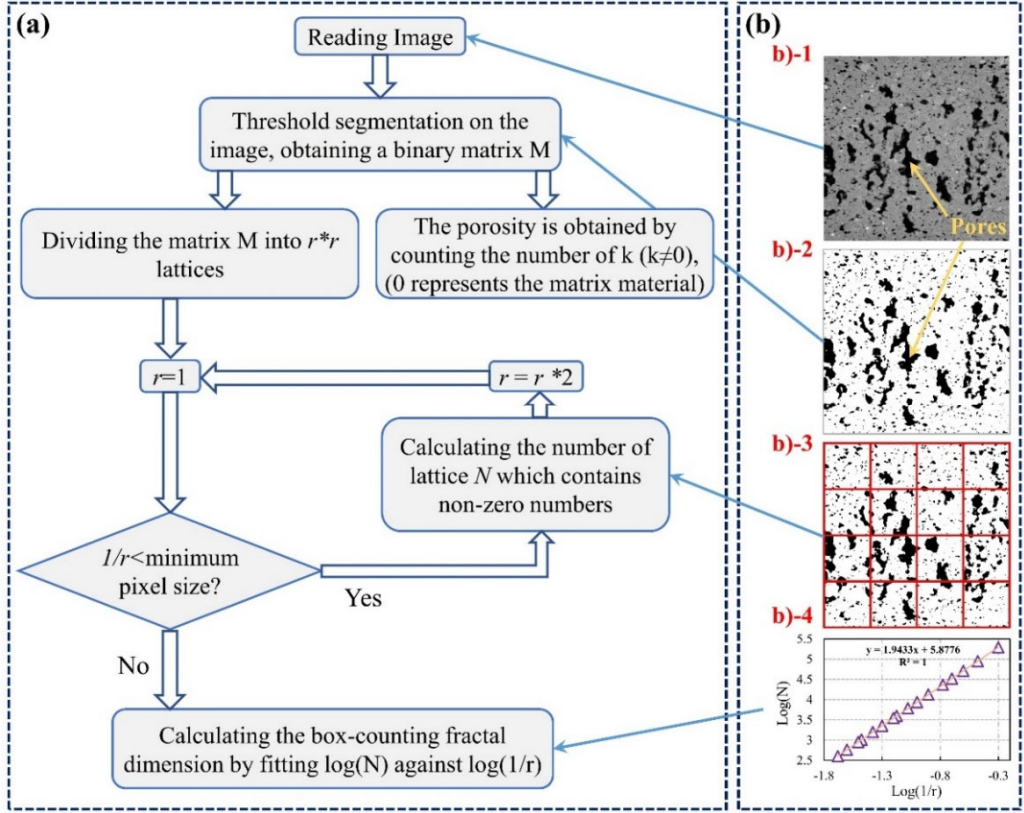

Fig. 4. Process of solving the plane porosity and box−counting dimension of a 2D slice

The fractal dimension can characterize the complexity of a structure from a multiscale perspective [35]. Mandelbrot et al. [36] pointed out that fractal dimension provides a new approach to describe the self-similarity and complexity of objects. Many calculation methods, such as box-counting dimension, triangulation and power spectrum, have been used to characterize the fractal characteristics of structures [37]. In this work, the box−counting dimension is used to evaluate the fractal characteristics of pores in 2D slices. For complex structures with autocorrelation, the box-counting dimension $D^b$ can be expressed as:

$$D^b = \log(N^k)/\log(1/k) \tag{3}$$

where $N^k$ is the minimum number of grids that overlay the surface of the complex structure with a square grid with side length $k$. The box−counting dimension indicates that the number of square grids $N^k$ increases with the decrease in side length $k$, and the calculation algorithm of the box−counting dimension on the 2D slice is shown in Fig.

4.

*2.7 Interlayer bonding strength of the printed structure*

The interlayer bond strength of 3D printed structure was measured by direct tensile loading test. Before the tensile loading test, the upper and lower ends of the specimen were cut and polished, and the total thickness of the specimen after polishing was approximately 20 mm. Then, the samples were bonded to the molds with high-strength epoxy glue, as shown in Fig. 5. In the tensile bond strength testing, tensile loading at a loading speed of 0.035 ± 0.015 MPa/s was applied to both the upper and lower ends of the specimen. The tensile bond strength is calculated through the ratio of the maximum tensile force to the effective bonding area.

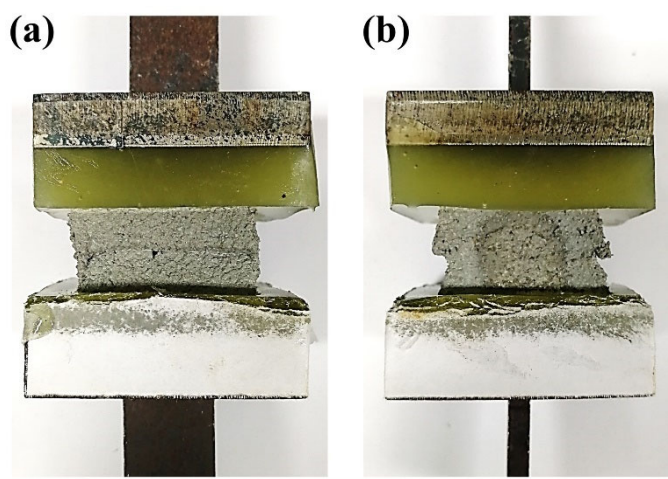

Fig. 5. Schematic of sample preparation for the interlayer bonding strength test.

## 3. Results and discussion

*3.1 Structural characterization of the PCs*

The FTIR spectra of HD-PC and FR-PC are reported in Fig. 6. The peak at approximately 3431 cm$^{-1}$ is the stretching vibration peak of hydroxyl (−OH), and the peak at approximately 2875 cm$^{-1}$ is the characteristic vibrational absorption peak of C−H. The stretching vibration absorption peak of carboxyl at 1730 cm$^{-1}$, the peak at approximately 1460 cm$^{-1}$ corresponds to the C−H bending vibration absorption peak of methylene (−CH$_2$−), the peak at approximately 1345 cm$^{-1}$ corresponds to the −OH bending vibration absorption peak of carboxyl, the peaks at approximately 1110 cm$^{-1}$ and 948 cm$^{-1}$ correspond to the C−O−C bending vibration absorption peak, and the peak at approximately 845 cm$^{-1}$ corresponds to the −CH$_2$− bending vibration absorption peak. The FTIR results show that the molecular structures of HD-PC and FR-PC polymers contain hydroxyl, carboxyl, ether, and other functional groups. In addition, from the FTIR spectra of FR-PC, it can be found that there is a significant peak at approximately 1583 cm$^{-1}$, which may be the characteristic absorption peak of acrylate moieties, corresponding to the single bond −COO$^-$ stretching vibration [38].

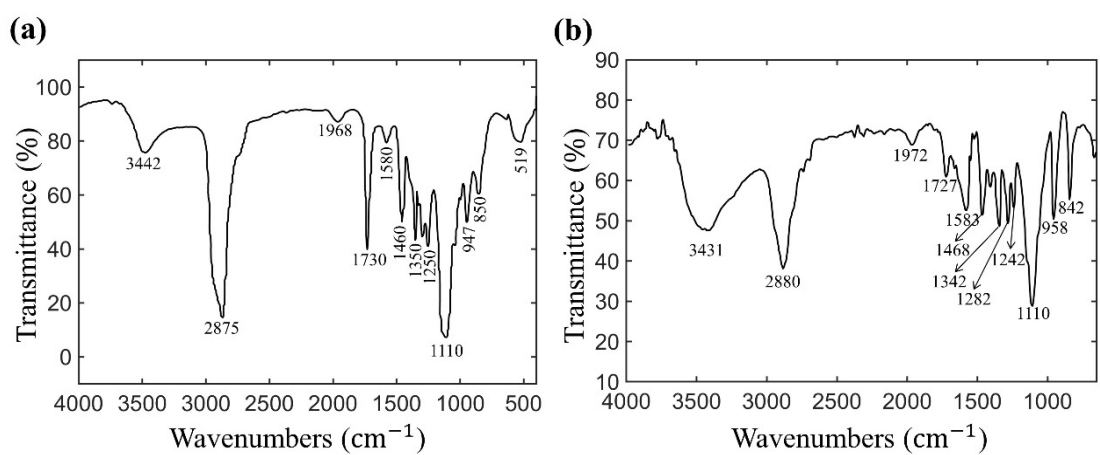

Fig. 6. FTIR spectra for (a) HD-PC and (b) FR-PC.

As shown in Fig. 7 and Table 4, the number-average molecular weights ($M_n$) of HD-PC are relatively close to 19930 g/mol, whereas the corresponding weight-average molecular weights ($M_w$) are 38993 g/mol. The weight−average molecular weight ($M_w$) and number−average molecular weight ($M_n$) of the FR-PC polymer are 15373 g/mol and 7362 g/mol in peak 1, respectively. In comparison with FR-PC, HD-PC exhibits a comparatively narrow molecular weight distribution.

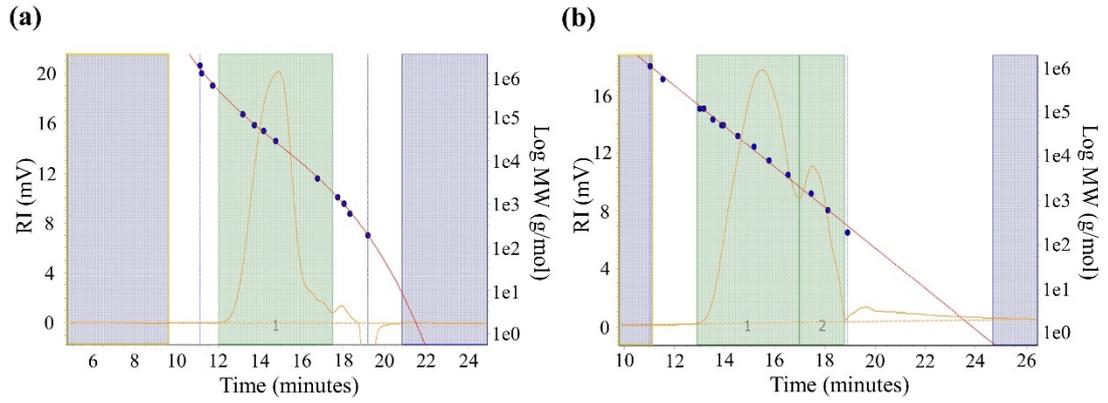

Fig. 7. Gel permeation chromatography (GPC) measurement results of HD-PC (a) and FR-PC (b).

Table 4. Molecular weights and their distributions in HD-PC and FR-PC.

| Polymer | Peak Results | $M_p$ (g/mol) | $M_n$ (g/mol) | $M_w$ (g/mol) | PDI ($M_w/M_n$) | Area (%) |
|---|---|---|---|---|---|---|
| HD-Pc | Peak 1 | 24532 | 19930 | 38993 | 1.956 | 100.00 |
| FR-PC | Peak 1 | 9764 | 7362 | 15373 | 2.088 | 76.02 |
|  | Peak 2 | 1198 | 860 | 1043 | 1.213 | 23.98 |

*3.2 Adsorption behaviors*

The adsorption behaviors of superplasticizers on the surface of cement particles may greatly change the physical and chemical properties of the solid–liquid interface and the interaction between particles through electrostatic repulsion and steric hindrance [39]. The –SO$_3$H, –OH and –COOH groups in the side chain result in the adsorption behavior of PCs [40, 41]. The adsorption amount of HD-PC and FR-PC on the cement particles is investigated, and the results are shown in Fig. 8.

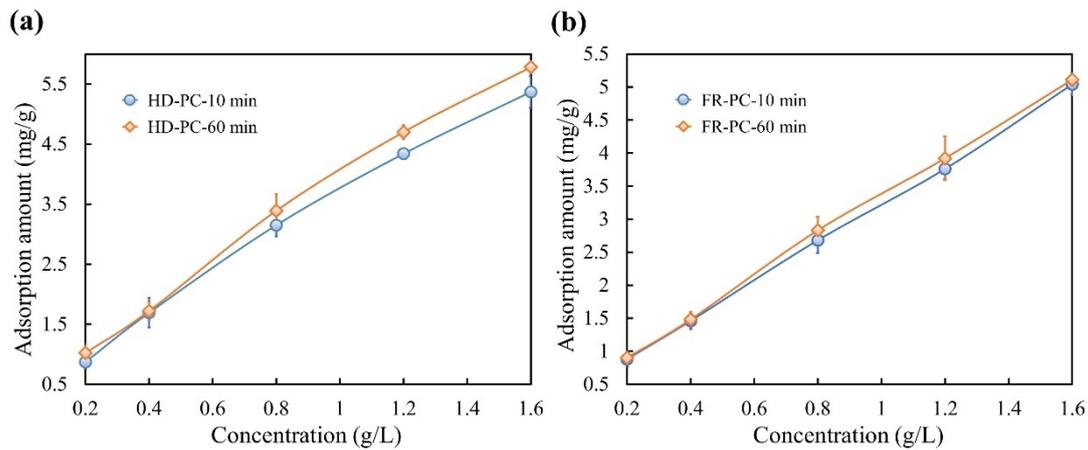

Fig. 8. Adsorption amount of HD-PC (a) and FR-PC (b) on the cement particles.

A higher adsorption amount is observed in HD-PC than FR-PC, indicating the stronger adsorption capacity of HD-PC than FR-PC. This may result in a higher flowability of the cement mixture with HD-PC polymer at the initial stage. Furthermore, from 10 min to 60 min, the adsorption amounts of HD-PC and FR-PC are both increased, because the deposition of hydration products on the surface of cement particles decreases the number of PC molecules in solution [42]. However, in comparison with FR-PC, higher increase ratios are observed in HD-PC. This result implied that with time, the adsorption amount of HD-PC would increase more than that of FR-PC. Under the alkaline condition of cement hydration, the ester groups of FR-PC can hydrolyze into carboxylate groups, which increases the concentration of carboxyl groups in solution [43].

*3.3 Fluidity*

The fluidity and its time-varying characteristics are investigated, and the results are shown in Fig. 9. Higher initial fluidity is observed in HD-PC than FR-PC. The initial fluidity of HD-PC is 6.5% higher than that of FR-PC. The higher initial fluidity of HD-PC could be attributed to the stronger adsorption capacity of HD-PC than FR-PC. A higher adsorption capacity of HD-PC would lead to a thicker water film. Consequently, a higher plasticizing effect and higher fluidity are observed in HD-PC at the beginning.

From 0 min to 30 min, the fluidity of HD-PC decreases by 23.1%, while that of FR-PC decreases by 14.4%. From 0 min to 60 min, the fluidity of HD-PC decreases by 37.1%, while that of FR-PC decreases by 22.3%. This indicates that the fluidity-retention ability of FR-PC is higher than that of HD-PC. The ester groups of FR-PC will gradually hydrolyze into adsorbable carboxylate ions in the alkaline environment of cement pastes; thus, more carboxyl groups could be adsorbed onto the cement particles to provide a secondary dispersion effect [44]. Therefore, the fluidity loss of cement paste with FR-PC is lower than that with HD-PC.

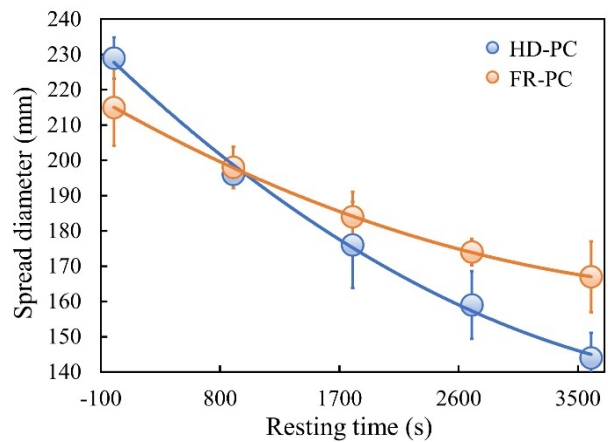

Fig. 9. Test results of fluidity of fresh paste.

*3.4 Rheological properties of cement mortars*

The test results of the thixotropic hysteresis loop and static yield stress are shown in Fig. 10(a, b). Compared with HD-PC, FR-PC significantly increases the thixotropic hysteresis loop area of cement mortar, and its thixotropic hysteresis loop area is 101.9% higher than that of HD-PC. This indicates that FR-PC exhibits stronger thixotropy than HD-PC. Thixotropy was characterized by two aspects, i.e., structural build-up of the suspension at rest and structural breakdown due to applied shearing [31, 45, 46]. In comparison with FR-PC, the higher adsorption capacity of HD-PC would lead to a higher electrostatic repulsive force, which would result in good dispersion of cement particles and hinder the flocculation process. Proper thixotropic behavior is an essential quality to ensure the printability and buildability of cementitious materials. As shown in Fig. 11, both the mixtures with HD-PC and FR-PC have good printability and buildability.

Fig. 10(b) reports the time–varying curve of the static yield stress, which can be fitted by a nonlinear equation [45, 47, 48], and the fitted results are summarized in Table 5. The increase in the static yield stress with time could be described by two successive gradients. The initial gradient of static yield shear stress evolution is depicted by $R_{thix}$, referring to the rapid rebuilding by reflocculation, which is a physical reaction that largely occurs in the first few hundred seconds after removal of applied energy [2, 13, 47]. The second gradient is depicted $A_{thix}$, which refers to a lower rate of strengthening, brought about a structuration process following reflocculation, which may be attributed to the hydration reactions of cement and formation of hydration products, typically in the thousands of seconds [47].

As reported in Table 5, a higher value of $R_{thix}$ is observed in FR-PC than in HD-PC. The reflocculation rate ($R_{thix}$) of FR-PC is 80.4% higher than that of HD-PC. The high dispersion performance of HD-PC limits the physical flocculation of cement particles at the beginning, which results in a lower reflocculation rate ($R_{thix}$). It is worth mentioning that $R_{thix}$ had the same trend as the thixotropy represented by the thixotropic hysteresis loop, which indicates that reflocculation rate may also be an indicator of appropriate thixotropic behavior for cement mortars. This was consistent with the research results of Kruger et al. [13, 47]. Furthermore, a higher value of $A_{thix}$ is observed in HD-PC than FR-PC. The structuration rate ($A_{thix}$) of HD-PC is 40.0% higher than that of FR-PC. This could be attributed to the lower fluidity-retention ability of HD-PC than FR-PC. With increasing resting time, the mixtures with HD-PC rapidly lose the dispersion capacity and electrostatic repulsive force, and the agglomeration structure accumulates and forms a network structure under the influence of van der Waals attraction forces. Simultaneously, the nucleation of early hydration products strengthens the connections between cement particles within the network. Therefore, the structuration rate of the mixture with HD-PC is higher than that with FR-PC.

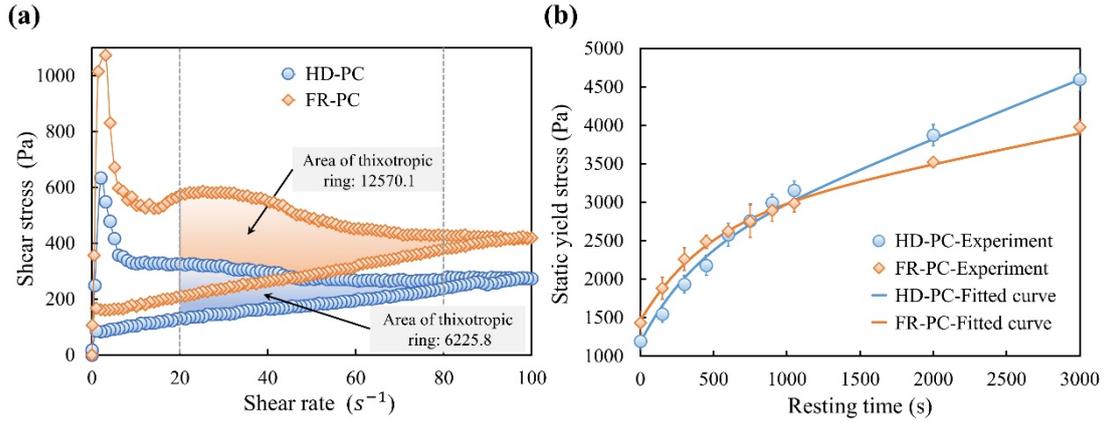

Fig. 10. Test results of the thixotropic ring area (a) and static yield stress (b) of fresh paste.

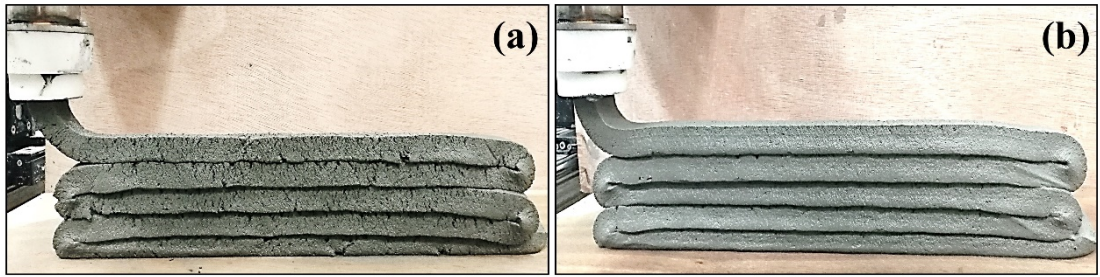

Fig. 11. Demonstration of 3D printability of mixtures: (a) HD-PC/T20, (b) FR-PC/T20.

Table 5. Thixotropic parameters for cement mortar with HD-PC polymer and FR-PC polymer.

| Types | $\tau_0$ (Pa) | $R_{thix}$ (Pa/s) | $t_{perc}$ (s) | $A_{thix}$ (Pa/s) |
| --- | --- | --- | --- | --- |
| HD-PC | 1194 | 1.53 | 543.2 | 0.71 |
| FR-PC | 1434 | 2.76 | 534.6 | 0.50 |

*3.5 Microstructure*

*3.5.1 Plane porosity and plane fractal dimension*

The printing ink with different rheological properties may lead to various microstructures at layer−to−layer interfaces, which would significantly affect the interlayer bonding strength of printed structures. In this work, plane porosity $\phi$ is used to characterize local air−void systems of 3D printed structures, and the plane fractal dimension $D^b$ is used to describe the complexity and irregularity of air−void systems [36]. Fig. 12-15 reports how the plane porosity $\phi$ and the plane fractal dimension $D^b$ change along the thickness of the printed structures, respectively. These values are obtained through digital image processing technology, as described in Section 2.6.2.

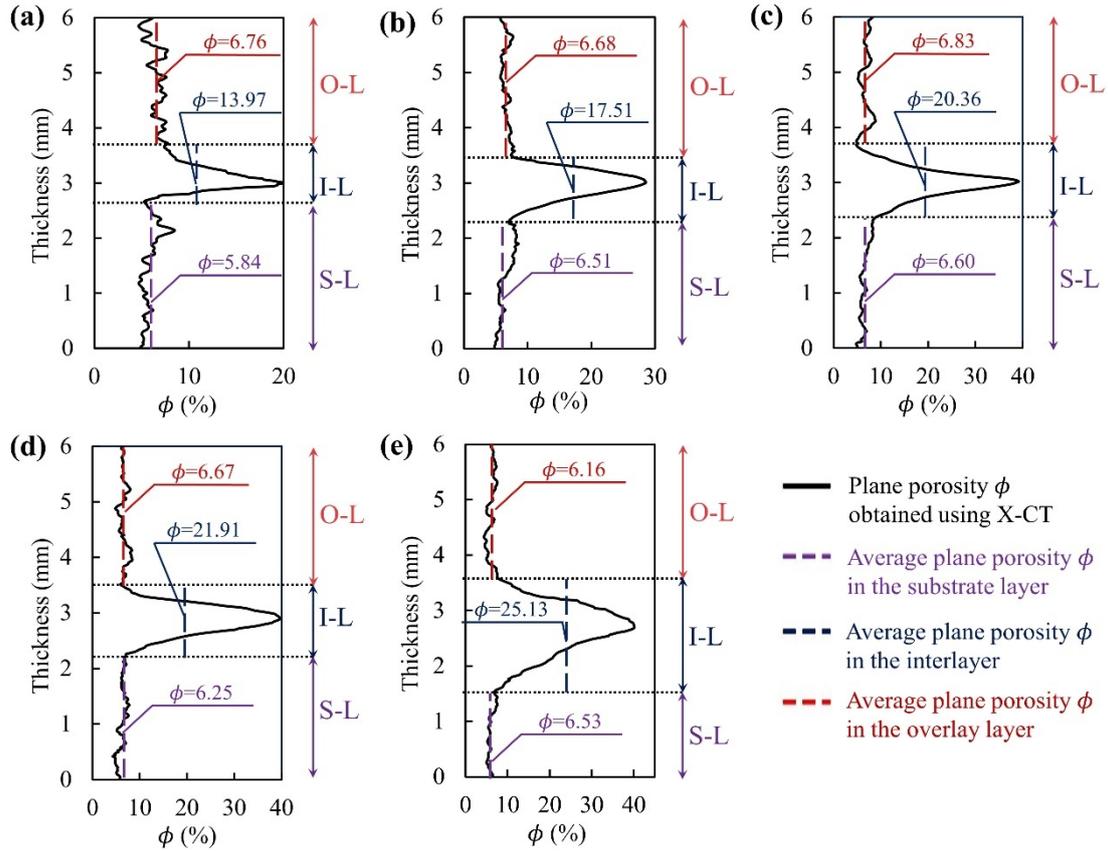

Fig. 12. Effect of time interval on the plane porosity $\phi$ of printed structures with HD-PC: (a) HD-PC/T20, (b) HD-PC/T900, (c) HD-PC/T1800, (d) HD-PC/T2700 and (e) HD-PC/T3600.

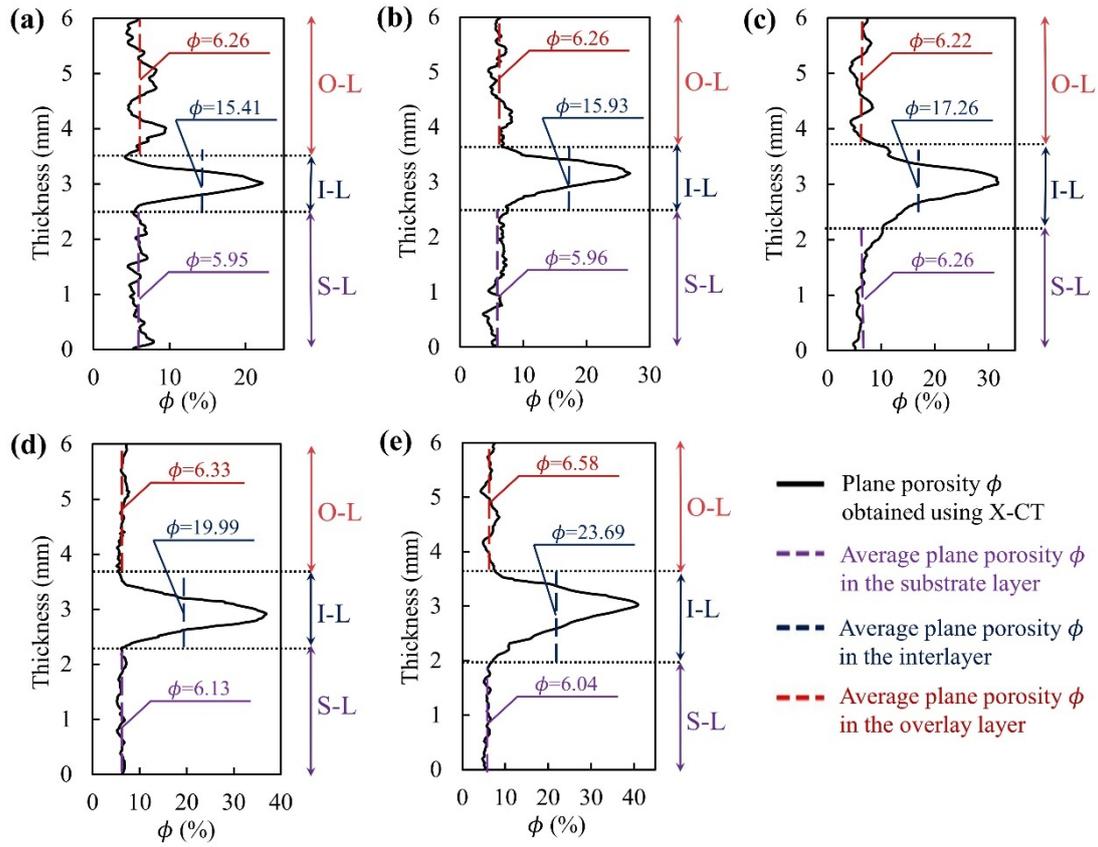

Fig. 13. Effect of time interval on the plane porosity $\phi$ of printed structures with FR-PC: (a) FR-PC/T20, (b) FR-PC/T900, (c) FR-PC/T1800, (d) FR-PC/T2700 and (e) FR-PC/T3600.

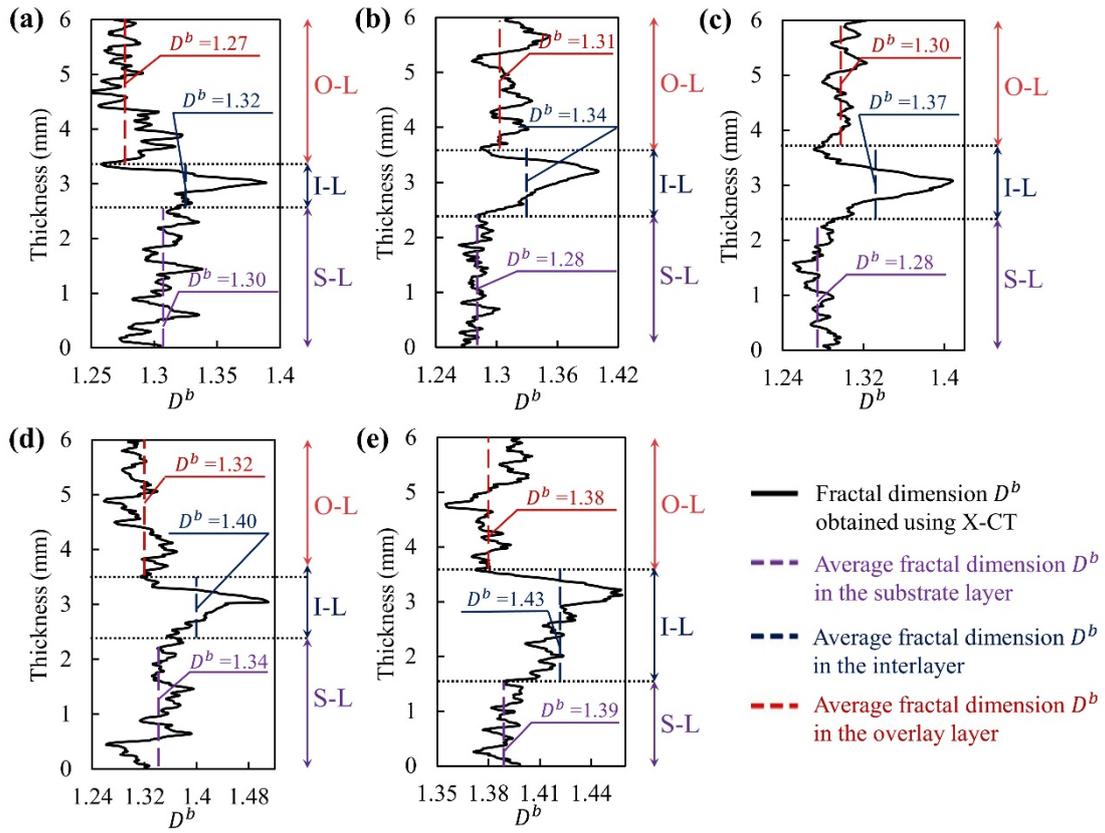

Fig. 14. Effect of time interval on the fractal dimension $D^b$ of printed structures with HD-PC: (a) HD-PC/T20, (b) HD-PC/T900, (c) HD-PC/T1800, (d) HD-PC/T2700 and (e) HD-PC/T3600.

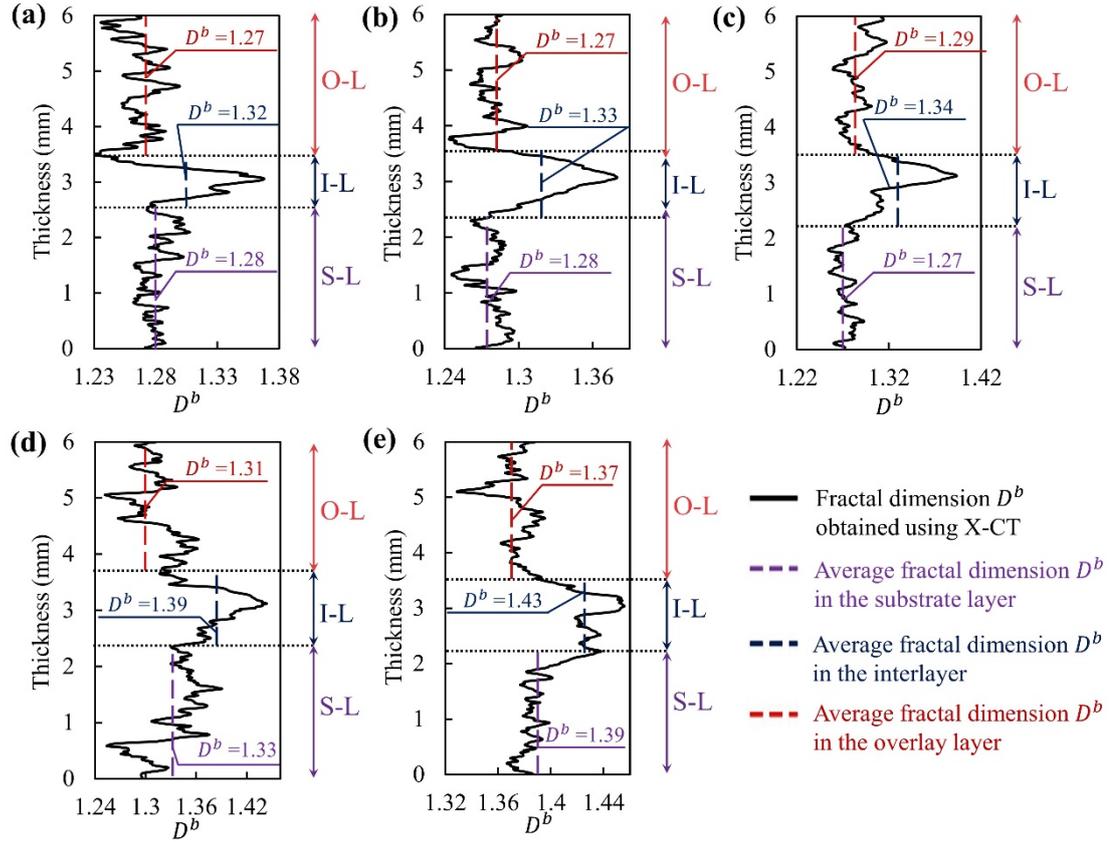

Fig. 15. Effect of time interval on the fractal dimension $D^b$ of printed structures with FR-PC: (a) FR-PC/T20, (b) FR-PC/T900, (c) FR-PC/T1800, (d) FR-PC/T2700 and (e) FR-PC/T3600.

As shown in Figs. 12-13, the plane porosity in the middle zone of the printed structures is higher than that in other areas, which is due to the lack of fusion between the adjacent extruded layers. Based on the stabilization of the plane porosity, where the interlayer zone begins and finishes is decided. With increasing of time interval, the average plane porosity of the interlayer zone (I-L) increases rapidly. For the samples with HD-PC, the average plane porosities $\phi$ of I-L with time intervals of 15 min, 30 min, 45 min and 60 min are 25.3%, 45.7%, 56.8% and 79.9% higher than that with a time interval of 20 s, respectively. For the samples with FR-PC, the average plane porosities $\phi$ of I-L with time intervals of 15 min, 30 min, 45 min and 60 min are 3.4%, 12.0%, 29.7% and 53.7 % higher than that with a time interval of 20 s, respectively. Extending the time interval will increase the surface stiffness of the substrate layer. A rough surface of the substrate layer may result in more difficulty in fusing between adjacent layers [17]. Extending the time interval promotes the evaporation of water between the two layers and increases the porosity of the interlayer zone. Furthermore, extending the time interval may also introduce 'wide' macropores in the interlayer zone. Many 'wide' macropores concentrated in the interlayer zone may increase the complexity and irregularity of the air−void systems. For the samples with HD-PC, the average fractal dimension $D^b$ of I-L with time intervals of 30 min and 60 min are 3.8% and 8.3% higher than that with a time interval of 20 s, respectively. For the samples with FR-PC, the average fractal dimension $D^b$ of I-L with a time interval of 60 min is 1.5% and 8.3% higher than that with a time interval of 20 s, respectively.

In addition, the rheological properties of fresh paste also significantly affect the microstructure of the interlayer zone. For a time interval of 20 s, the average plane porosity $\phi$ of I-L with FR-PC is 10.3% higher than that with HD-PC. This may be attributed to the higher initial fluidity of HD-PC than FR-PC. The good wettability and fluidity of the printing ink are expected to successfully fill and permeate into the pores and cracks in the surface of the previously deposited layer, which increases the compactness and reduces the average plane porosity $\phi$ of I-L. For a time interval greater than 20 s, the average plane porosity $\phi$ of I-L with HD-PC is higher than that with the FR-PC polymer. For a time interval of 15 min, the average plane porosity $\phi$ of I-L with FR-PC is 9.0% lower than that with HD-PC. For a time interval of 30 min, the average plane porosity $\phi$ of I-L with FR-PC is 15.2% lower than that with HD-PC. This may be attributed to the lower fluidity-retention ability of HD-PC than FR-PC. If the mixture with HD-PC is extruded out of the nozzle, it will lose plasticity and harden quickly, which is beneficial to the buildability, but results an increase in the stiffness difference and porosity between the two layers.

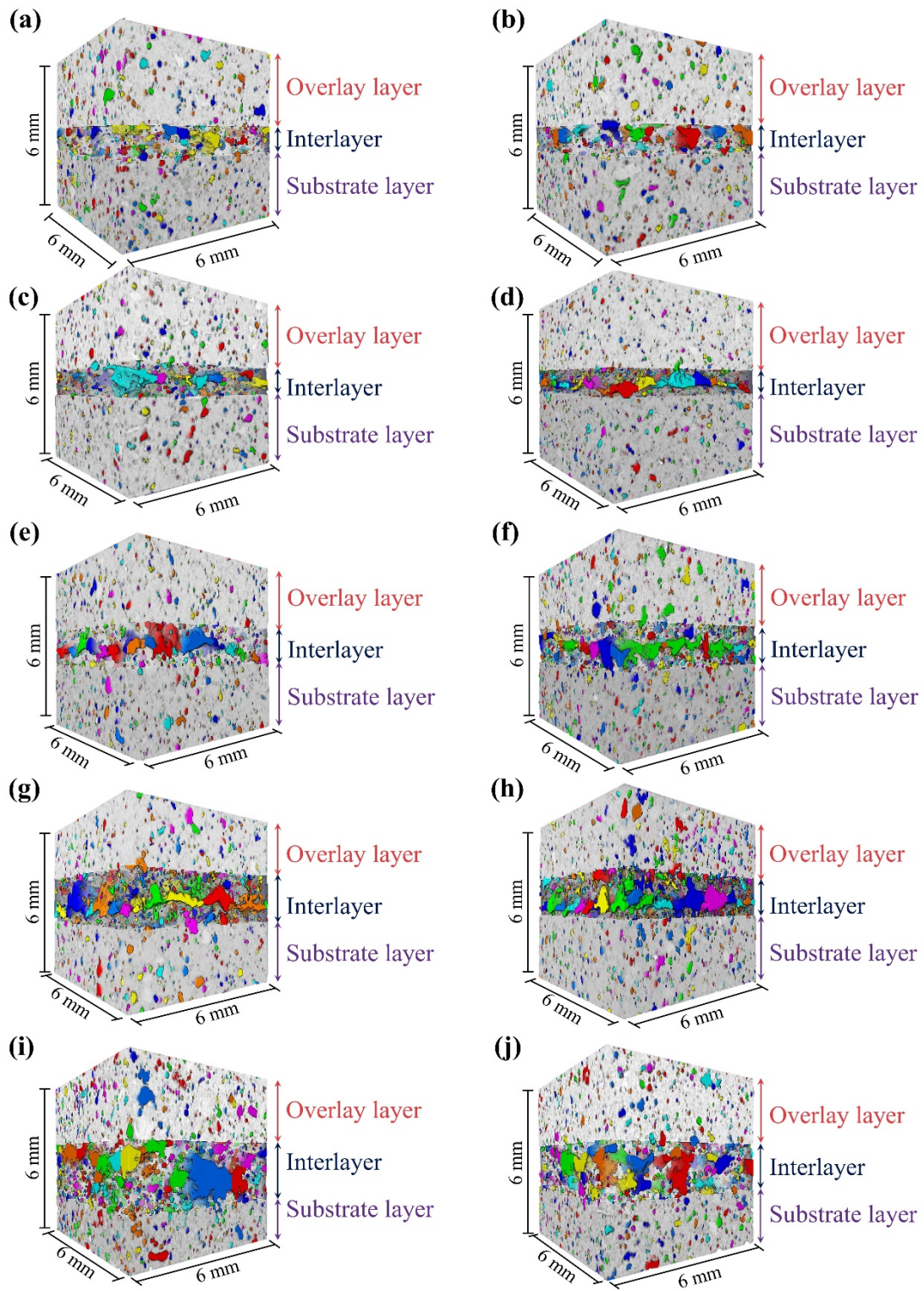

Fig. 16. Reconstructions of the 3D microstructure of 3D printed structures, which were obtained on the basis of tests using the micro−CT method for samples: (a) HD-PC/T20, (b) FR-PC/T20, (c) HD-PC/T900, (d) FR-PC/T900, (e) HD-PC/T1800, (f) FR-PC/T1800, (g) HD-PC/T2700, (h) FR-PC/T2700, (i) HD-PC/T3600, (j) FR-PC/T3600.

Fig. 16 shows the reconstruction of the 3D microstructure of double−layer printed structures based on the CT slice image using the X−CT test. Qualitative visual assessment of these 3D models shows that there is a significant interlayer zone between the substrate layer and overlay layer. There is a significant difference in the microstructure of the substrate layer, overlay layer and interlayer zone of the printed structure. Many small unconnected pores are contained in the substrate layer and overlay layer, as shown in Fig. 16. Each color in Fig. 16 represents an independent pore structure after image segmentation. Generally, the compactness of the substrate layer is higher than that of the overlay layer because the nozzle and extruded ink may exert pressure on the substrate layer during the extrusion-deposition process. Many unfilled areas and 'wide' macropores are concentrated on the interlayer zone of printed structures, which provides a channel for chloride or carbon dioxide ion ingress and is the possible reason for the weak interlayer bonding strength and poor durability properties of printed structures [24].

*3.5.2 Maximum plane porosity and plane fractal dimension*

The maximum plane porosity $\phi_{max}$ and maximum plane fractal dimension $D^b_{max}$ in the interlayer zone are therefore investigated and plotted in Fig. 17. Both the time interval and superplasticizer types significantly affect the maximum value of plane porosity $\phi_{max}$ and fractal dimension $D^b_{max}$ in the interlayer zone.

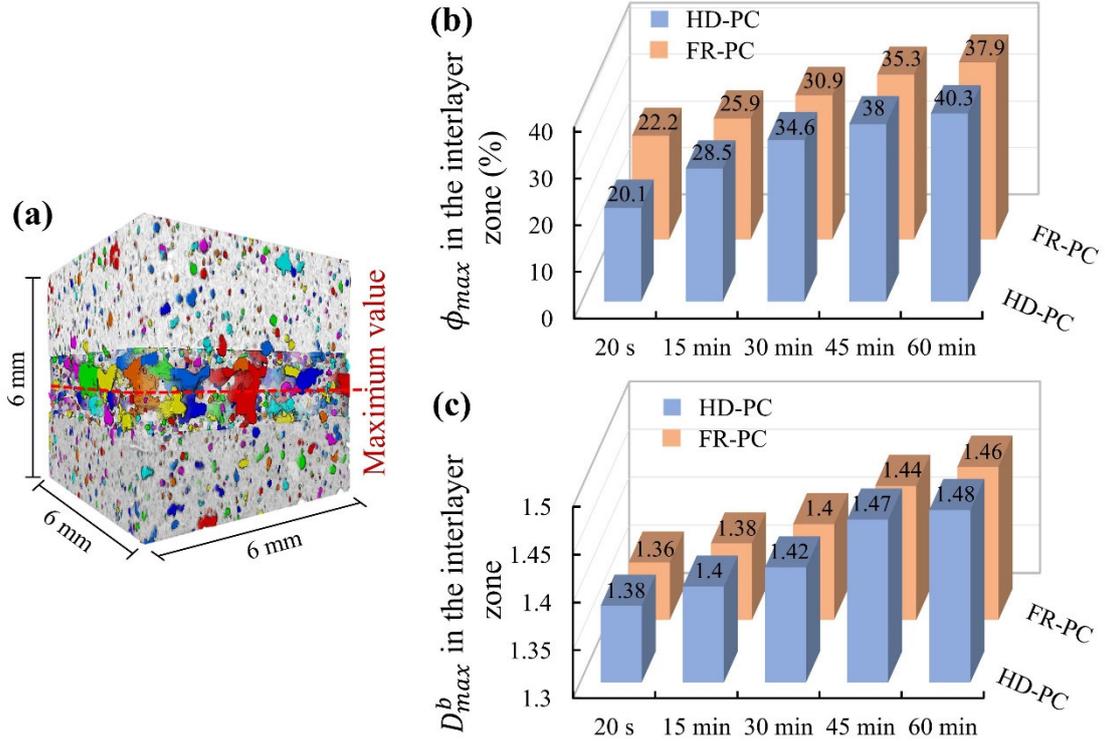

Fig. 17. Interlayer zone in the printed structures (a); maximum plane porosity in the interlayer zone (b); maximum plane fractal dimension in the interlayer zone (c).

As shown in Fig. 17(b, c), the values of $\phi_{max}$ and $D^b_{max}$ increase with the time interval. The values of $\phi_{max}$ and $D^b_{max}$ with a time interval of 20 s are lowest. Extending the time interval would result in a significant increase in these values. For the samples with HD-PC, the values of $\phi_{max}$ with time intervals of 15 min, 30 min, 45 min and 60 min are 41.8%, 72.1%, 89.1% and 100.5% higher than those with a time interval of 20 s, respectively, and the values of $D^b_{max}$ with time intervals of 15 min, 30 min, 45 min and 60 min are 1.4%, 2.9%, 6.5% and 7.2% higher than those with a time interval of 20 s, respectively. For the samples with FR-PC, the values of $\phi_{max}$ with time intervals of 15 min, 30 min, 45 min and 60 min are 16.7%, 39.2%, 59.0% and 70.7% higher than those with a time interval of 20 s, respectively, and the values of $D^b_{max}$ with time intervals of 15 min, 30 min, 45 min and 60 min are 1.5%, 2.9%, 5.9% and 7.4% higher than those with a time interval of 20 s, respectively. This indicates that extending time interval will increase the porosity of I-L and exert a negative impact on the fusion of adjacent layers.

Furthermore, the printed structures with FR-PC always have lower values of $\phi_{max}$ than those with HD-PC, except for the sample with a time interval of 20 s. For a time interval of 15 min, the value of $\phi_{max}$ with FR-PC is 9.1% lower than that with HD-PC. For a time interval of 30 min, the value of $\phi_{max}$ with FR-PC is 10.7% lower than that with HD-PC. For a time interval of 60 min, the value of $\phi_{max}$ with FR-PC is 6.0% lower than that with HD-PC. This indicates that FR-PC has a positive effect on the fusion of two layers and reduces the maximum porosity of the interlayer zone. For a time interval of 20 s, the value of $\phi_{max}$ with FR-PC is 10.4% higher than that with HD-PC. This may be attributed to the lower initial fluidity of FR-PC. Low fluidity reduces the filling effect and penetration behavior between adjacent layers, which will increase the porosity of the interlayer zone of printed structures. As shown in Fig. 17(a), the growth rate of the maximum plane porosity with FR-PC is lower than that with HD-PC. From 20 s to 15 min, the maximum plane porosity $\phi_{max}$ of I-L with HD-PC increases by 41.8%, while that with FR-PC increases by 16.7%. From 20 s to 30 min, the maximum plane porosity $\phi_{max}$ of I-L with HD-PC increases by 72.1%, while that with FR-PC increases by 39.2%. From 20 s to 60 min, the maximum plane porosity $\phi_{max}$ of I-L with HD-PC increases by 100.5%, while that with FR-PC increases by 70.7%. This indicates that the fluidity-retention ability of FR-PC may reduce the negative effect caused by extending the time interval and slow the growth rate of porosity in the interlayer zone with increasing time interval.

*3.6 Interlayer bonding strength*

Fig. 18 reports the interlayer bonding strength of 3D printed structures using uniaxial tensile test. The interlayer bonding strength of printed structures with time interval of 20 s is the largest. They are 1.93 MPa and 1.74 MPa for the printed structures with HD-PC and FR-PC, respectively. Extending the time interval will exert a negative effect on the interlayer bonding strength of the printed structure. The interlayer bonding strength decreases rapidly with increasing time interval. For the printed structures with HD-PC, the interlayer bonding strengths with time intervals of 15 min, 30 min, 45 min and 60 min are 32.1%, 60.1%, 73.6% and 79.8% lower than that with a time interval of 20 s, respectively. For the printed structures with FR-PC, the interlayer bonding strengths with time intervals of 15 min, 30 min, 45 min and 60 min are 12.6%, 40.2%, 53.4% and 71.3% lower than those with a time interval of 20 s, respectively. This phenomenon may be explained by the fact that a large printing time interval increases the modulus/stiffness difference between two layers and the evaporation of water in the interlayer zone, increases the porosity of the interlayer zone, and reduces the contact area and interlayer bonding strength of the printed structure.

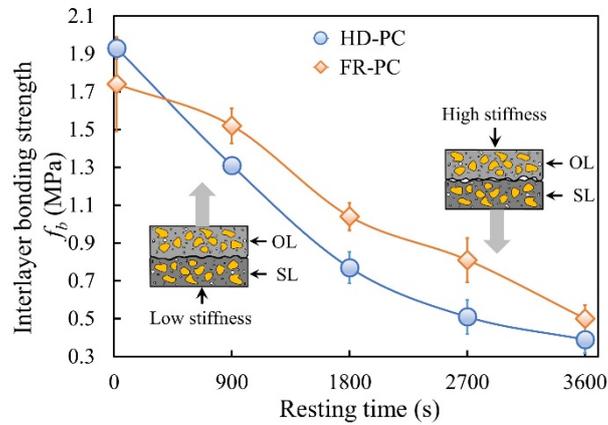

Fig. 18. Interlayer bonding strength of 3D printed structures.

Furthermore, the printed structures with FR-PC always have higher interlayer bonding strength than those with HD-PC, except for the sample with a time interval of 20 s. For a time interval of 20 s, the interlayer bonding strength of the mixture with FR-PC is 9.8% lower than that with HD-PC. For a time interval of 15 min, the interlayer bonding strength with FR-PC is 16.0% higher than that with HD-PC. For a time interval of 30 min, the interlayer bonding strength with FR-PC is 35.1% higher than that with HD-PC. For a time interval of 60 min, the interlayer bonding strength with FR-PC is 28.2% higher than that with HD-PC. This indicates that FR-PC has a positive effect on the interlayer bonding of the printed structure, especially for long time intervals.

Furthermore, by comparing the results illustrated in Fig. 17 and Fig. 18, the change trend of the maximum plane porosity $\phi_{max}$ of I-L is opposite to that of the interlayer bonding strength. The larger the value of $\phi_{max}$ is, the smaller the interlayer bonding strength. This phenomenon is not difficult to understand; the high porosity of interlayer zone would reduce the contact area between the two layers and thus result in a low interlayer bonding strength. For a quasi−brittle 3D printed cement mortar, the interlayer bonding strength is closely related to the air−void systems within the interlayer zone [49, 50].

*3.7 Relationship between air-void systems and interlayer bonding strength*

The potential correlation between the microstructure of the interlayer zone and interlayer bonding strength is investigated. Fig. 19 reports the relations between the average plane porosity $\phi$, average fractal dimension $D^b$, maximum plane porosity $\phi_{max}$, maximum fractal dimension $D^b_{max}$ and interlayer bonding strength of 3D printed structures.

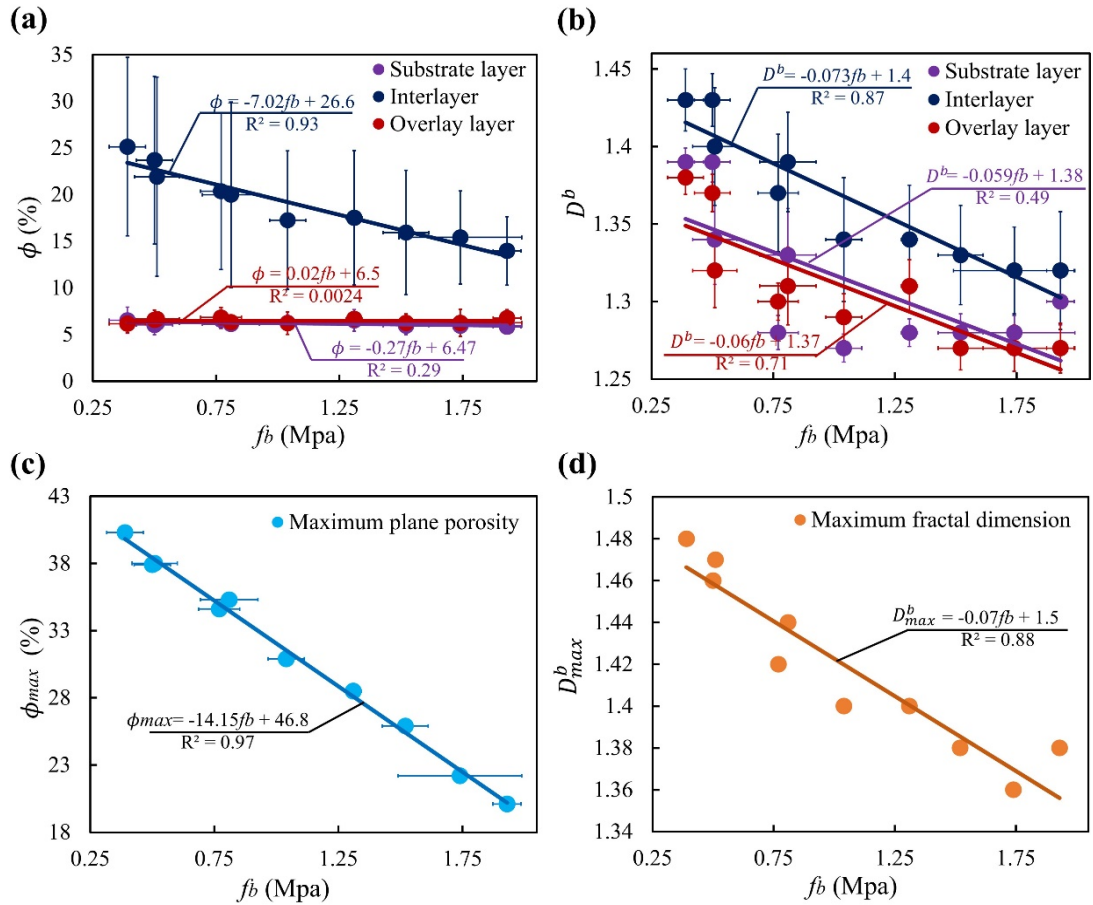

Fig. 19. Relationships between the average plane porosity $\phi$ (a), average plane fractal dimension $D^b$ (b), maximum plane porosity $\phi_{max}$ (c), maximum plane fractal dimension $D^b_{max}$ (d) and the interlayer bonding strength of the 3D printed structure.

According to the principle of minimum energy consumption of fracture mechanics, the crack propagation path always follows the surface with the weakest bonding force [51]. For the 3D printed structures, the weak bonding force mainly concentrated on the interlayer zone, due to the higher plane porosity in the interlayer zone than in the other zones, as shown in Figs. 12-13. This is a significant negative linear correlation between the plane porosity of the interlayer zone and interlayer bonding strength, as shown in Fig .19(a, c). There is a stronger relationship between the maximum plane porosity $\phi_{max}$ and interlayer bonding strength ($R^2 = 0.97$) than between the average plane porosity of I-L and interlayer bonding strength ($R^2 = 0.93$) because the interface with maximum plane porosity obtains the smallest contact area and the weakest interlayer adhesion. In the tensile loading test of printed structures, the microcracks propagate mainly along the interlayer zone in the tensile process, especially along the slice with the maximum plane porosity. Compared with the average plane porosity of I-L, the average plane porosities of substrate layer (S-L) and overlay layer (O-L) have little effect on the interlayer bonding strength because the tensile failure mainly occurs in the interlayer zone.

Furthermore, as shown in Fig. 19(b, d), there is no statistically significant correlation between the plane fractal dimension and the interlayer bonding strength. This result may be attributed to the possible errors by the image-based fractal calculation method. The 2D fractal dimensions calculated from the box-counting method truly reflect the irregularities of pore structure on the 2D images, but the fractal dimension with 2D images often underestimates the complexity of the actual 3D structure [52].

## 4. Conclusions

In this study, the structural characterization, adsorption behaviors and time varying characteristics of rheological properties of two types of PC (a high dispersion polycarboxylate superplasticizer HD-PC and a fluidity−retaining polycarboxylate superplasticizer FR-PC) were investigated. Furthermore, the correlation of printing time interval, time varying characteristics of rheological properties, microstructure and interlayer bonding strength was also studied. The following results are obtained:

(1) FR-PC has a lower initial fluidity than HD-PC, but its fluidity-retention ability is higher than that of HD-PC because the amide monomers can gradually hydrolyze into adsorbable carboxylate ions in the alkaline environment of cement pastes.

(2) The thixotropic hysteresis loop area of FR-PC is 101.9% higher than that of HD-PC. The reflocculation rate ($R_{thix}$) of FR-PC is 80.4% higher than that of HD-PC, while the structuration rate ($A_{thix}$) is 29.6% lower than that of HD-PC.

(3) Extending the time interval will exert a negative effect on the interlayer bonding strength of the printed structure. Extending the time interval increases the porosity of the interlayer zone and reduces the contact area and interlayer bonding strength of the printed structure. Compared with the plane porosity of the substrate layer and overlay layer, the maximum plane porosity and average plane porosity of the interlayer zone are more related to the interlayer bonding strength.

(4) The FR-PC polymer has a positive effect on the interlayer bonding of the printed structure, especially for the long time intervals. For a time interval of 60 min, the interlayer bonding strength of the mixture with FR-PC is 28.2% higher than that with HD-PC. Furthermore, high fluidity−retaining FR-PC may reduce the negative effect caused by extending the time interval on the interlayer bonding. From 20 s to 30 min, the maximum plane porosity $\phi_{max}$ in the interlayer zone with HD-PC increases by 50.0%, while that with FR-PC increases by 14.1%, the interlayer bonding strength with HD-PC decreases by 50.0 %, while that with FR-PC decreases by 14.1%.

# Appendix A: Adsorption amount of HD-PC and FR-PC on the cement particles

Table A1. The results of the adsorption amount of HD-PC and FR-PC on the cement particles.

| Mixture | Concentration (g/L) | Adsorption amount (mg/g) | | | Mean value (mg/g) | Standard deviation σ (mg/g) |
|---|---|---|---|---|---|---|
| | | Test result (mg/g) | | | | |
| HD-PC-10 min | 0.2 | 0.78 | 0.81 | 1.02 | 0.87 | 0.107 |
| | 0.4 | 1.43 | 1.62 | 2.02 | 1.69 | 0.246 |
| | 0.8 | 2.99 | 3.05 | 3.41 | 3.15 | 0.185 |
| | 1.2 | 4.25 | 4.31 | 4.46 | 4.34 | 0.088 |
| | 1.6 | 5.16 | 5.21 | 5.74 | 5.37 | 0.262 |
| SR-PC-10 min | 0.2 | 0.79 | 0.81 | 1.04 | 0.88 | 0.113 |
| | 0.4 | 1.34 | 1.41 | 1.63 | 1.46 | 0.124 |
| | 0.8 | 2.51 | 2.58 | 2.95 | 2.68 | 0.193 |
| | 1.2 | 3.59 | 3.72 | 3.97 | 3.76 | 0.158 |
| | 1.6 | 4.86 | 5.01 | 5.25 | 5.04 | 0.161 |
| HD-PC-60 min | 0.2 | 0.86 | 0.98 | 1.22 | 1.02 | 0.150 |
| | 0.4 | 1.59 | 1.64 | 1.93 | 1.72 | 0.150 |
| | 0.8 | 3.18 | 3.21 | 3.78 | 3.39 | 0.276 |
| | 1.2 | 4.61 | 4.62 | 4.87 | 4.7 | 0.120 |
| | 1.6 | 5.74 | 5.71 | 5.89 | 5.78 | 0.079 |
| SR-PC-60 min | 0.2 | 0.82 | 0.85 | 1.03 | 0.9 | 0.093 |
| | 0.4 | 1.35 | 1.46 | 1.63 | 1.48 | 0.115 |
| | 0.8 | 2.63 | 2.75 | 3.11 | 2.83 | 0.204 |
| | 1.2 | 3.52 | 3.91 | 4.33 | 3.92 | 0.331 |
| | 1.6 | 5.06 | 5.09 | 5.18 | 5.11 | 0.051 |

# Appendix B: Fluidity of fresh pastes with HD-PC or FR-PC

Table B1. The results of the fluidity of fresh pastes with HD-PC or FR-PC.

| Mixture | Time (s) | Spread diameter (mm) | | | Mean value (mm) | Standard deviation σ (mm) |
|---|---|---|---|---|---|---|
| | | Test result (mm) | | | | |
| HD-PC | 0 | 221 | 231 | 235 | 229 | 5.89 |
| | 900 | 194 | 195 | 199 | 196 | 2.16 |
| | 1800 | 165 | 170 | 193 | 176 | 12.19 |
| | 2700 | 146 | 162 | 169 | 159 | 9.63 |
| | 3600 | 134 | 149 | 149 | 144 | 7.07 |

|  | 0 | 205 | 210 | 230 | 215 | 10.8 |
|---|---|---|---|---|---|---|
|  | 900 | 190 | 200 | 204 | 198 | 5.89 |
| SR-PC-10 min | 1800 | 178 | 180 | 194 | 184 | 7.12 |
|  | 2700 | 169 | 178 | 175 | 174 | 3.74 |
|  | 3600 | 158 | 162 | 181 | 167 | 10.03 |

**Appendix C: Static yield stress of fresh pastes with HD-PC and FR-PC**

Table C1. The results of the Static yield stress of fresh pastes with HD-PC and FR-PC.

| Mixture | Time (s) | Static yield stress (MPa) | | | Mean value (Mpa) | Standard deviation σ (MPa) |
|---|---|---|---|---|---|---|
|  |  | Test result (Mpa) | | | | |
| HD-PC | 0 | 1084.54 | 1145.62 | 1352.56 | 1194.24 | 114.6927 |
|  | 150 | 1426.85 | 1526.42 | 1687.01 | 1546.76 | 107.1793 |
|  | 300 | 1808.61 | 1915.56 | 2081.58 | 1935.25 | 112.3059 |
|  | 450 | 2056.28 | 2124.17 | 2359.13 | 2179.86 | 129.7576 |
|  | 600 | 2438.39 | 2509.64 | 2783.78 | 2577.27 | 148.8935 |
|  | 750 | 2565.62 | 2654.27 | 3063.56 | 2761.15 | 216.8772 |
|  | 900 | 2857.67 | 2985.42 | 3134.17 | 2992.42 | 112.9891 |
|  | 1050 | 3009.53 | 3149.27 | 3307.4 | 3155.4 | 121.6821 |
|  | 2000 | 3746.49 | 3805.53 | 4068.12 | 3873.38 | 139.7955 |
|  | 3000 | 4459.36 | 4528.27 | 4808.05 | 4598.56 | 150.7795 |
| FR-PC | 0 | 1305.62 | 1358.51 | 1639.91 | 1434.68 | 146.7171 |
|  | 150 | 1702.92 | 1895.24 | 2052.67 | 1883.61 | 143.0215 |
|  | 300 | 2051.94 | 2352.46 | 2379.44 | 2261.28 | 148.435 |
|  | 450 | 2386.76 | 2481.53 | 2598.92 | 2489.07 | 86.7779 |
|  | 600 | 2586.49 | 2603.85 | 2682.35 | 2624.23 | 41.70366 |
|  | 750 | 2561.46 | 2646.98 | 3049.03 | 2752.49 | 212.5722 |
|  | 900 | 2786.86 | 2804.61 | 3092.84 | 2894.77 | 140.244 |
|  | 1050 | 2894.56 | 2918.49 | 3149.18 | 2987.41 | 114.8051 |
|  | 2000 | 3486.19 | 3504.28 | 3575.08 | 3521.85 | 38.35698 |
|  | 3000 | 3897.25 | 3915.57 | 4119.47 | 3977.43 | 100.7155 |

**Appendix D: Interlayer bonding strength of 3D printed structures**

Table D1. The results of the interlayer bonding strength $f_b$ and their statistical characteristics.

| Mixtures | No. | Interlayer bonding strength $f_b$ (MPa) | Mean value $\overline{f_b}$ (MPa) | Standard deviation $\sigma$ (MPa) |
|---|---|---|---|---|
| HD-PC/T20 | 1 | 1.97 | 1.93 | 0.057 |
| | 2 | 1.85 | | |
| | 3 | 1.97 | | |
| HD-PC/T900 | 4 | 1.35 | 1.31 | 0.033 |
| | 5 | 1.31 | | |
| | 6 | 1.27 | | |
| HD-PC/T1800 | 7 | 0.88 | 0.77 | 0.083 |
| | 8 | 0.75 | | |
| | 9 | 0.68 | | |
| HD-PC/T2700 | 10 | 0.61 | 0.51 | 0.091 |
| | 11 | 0.53 | | |
| | 12 | 0.39 | | |
| HD-PC/T3600 | 13 | 0.48 | 0.39 | 0.073 |
| | 14 | 0.39 | | |
| | 15 | 0.3 | | |
| FR-PC/T20 | 16 | 1.43 | 1.74 | 0.249 |
| | 17 | 1.75 | | |
| | 18 | 2.04 | | |
| FR-PC/T900 | 19 | 1.47 | 1.52 | 0.093 |
| | 20 | 1.44 | | |
| | 21 | 1.65 | | |
| FR-PC/T1800 | 22 | 1.04 | 1.04 | 0.073 |
| | 23 | 0.95 | | |
| | 24 | 1.13 | | |
| FR-PC/T2700 | 25 | 0.76 | 0.81 | 0.116 |
| | 26 | 0.7 | | |
| | 27 | 0.97 | | |
| FR-PC/T3600 | 28 | 0.5 | 0.5 | 0.073 |
| | 29 | 0.41 | | |
| | 30 | 0.59 | | |

**CRediT authorship contribution statement**

Tinghong Pan: Conceptualization, Methodology, Writing − Original draft, Writing – Review and Editing, Supervision. Yaqing Jiang: Conceptualization, Methodology, Investigation, Formal analysis, Review and Editing. Xuping Ji: Conceptualization, Methodology, Investigation, Formal analysis.

**Declaration of competing interest**

The authors declare that they have no known competing financial interests or personal relationships that could have appeared to influence the work reported in this paper.

**Acknowledgments**

This work was supported by the National Natural Science Foundation of China [grant numbers 51738003, 11772120].